\documentstyle[11pt,paspconf]{article}

\begin{document}

\title{Building up the Stellar Halo of the Milky Way}

\author{Amina Helmi}
\affil{Sterrewacht Leiden, Postbus 9513, 2300 RA Leiden, 
The Netherlands}
\author{Simon D.M. White}
\affil{Max-Planck-Institut f\"ur Astrophysik, Karl-Schwarzschild-Str. 
1, 85740 Garching bei M\"unchen, Germany}

\begin{abstract}
We study numerical simulations of satellite galaxy disruption
in a potential resembling that of the Milky Way. Our goal is
to assess whether a merger origin for the stellar halo would 
leave observable fossil structure in the phase-space distribution 
of nearby stars. We show how mixing of disrupted satellites can
be quantified using a coarse-grained entropy. Although after 10
billion years few obvious asymmetries remain in the distribution of 
particles in configuration space, strong correlations are still present
in velocity space. 
We briefly describe how we can understand these effects based on the 
conservation of fine-grained phase-space 
density in an action-angle formalism. We also discuss the 
implications of our results on the known properties of the stellar halo.
\end{abstract}

\keywords{Galaxy: halo, formation, dynamics -- 
galaxies: formation, halos, interactions}

\section{Introduction}
Currently popular theories of structure formation in the Universe
postulate that structure grows through the amplification by gravitational
forces of initially small density fluctuations (see White 1996 for a review). 
Depending on the  characteristics of the spectrum of fluctuations, 
small objects can be the first to collapse; they then merge to 
form progressively larger systems giving rise to the complex 
structure we observe today. In this hierarchical structure formation scenario
our Galaxy, as a typical galaxy, should also have been formed in part 
by merging and accretion of smaller galaxies, or `building blocks'. 
These events should be imprinted in some of its present-day components, 
presumably as residual structure.
For example, when a galaxy is disrupted it leaves trails of stars along
its orbit. These could be superposed in a spheroidal component such
as a stellar halo. In fact, numerous observations suggest substructure in the
halo of the Galaxy (see Majewski in this volume). 
In this paper, we attempt to describe what the signatures of 
different accretion events should be if indeed 
our Galaxy formed as envisaged in current theories. 
Should this merging history be observed in star counts, kinematic or 
abundance surveys of the Galaxy? How prominent or
not would these substructures be? How well-mixed are the stars that
made up these progenitors? 
What can we say about the properties of the accreted satellites 
from the observations we have today?  

\section{Simulations and Results}

To tackle the questions we just posed we carry
out N-body simulations of accretion of satellite galaxies, where we
represent the Milky Way by a fixed,
rigid potential and the satellite by a collection of
particles. The self-gravity of the satellite is modelled by 
a monopole term as in White (1983).

The Galactic potential is represented by two components:
a disk described by a Miyamoto-Nagai potential,
\begin{equation}
\label{eq:disk}
\Phi_{\rm disk} = - \frac{G M_{\rm disk}}{\sqrt{R^2 + (a + 
\sqrt{z^2 + b^2})^2}},
\end{equation}
where $M_{\rm disk} = 10^{11}\, {\rm M_{\odot}}$, $a = 6.5\, {\rm kpc}$, 
$b = 0.26 \,{\rm kpc}$, and a dark halo with a logarithmic potential,
\begin{equation}
\label{eq:halo}
\Phi_{\rm halo} = v^2_{\rm halo} \ln (r^2 + d^2),
\end{equation}
with $d = 12 \,{\rm kpc}$ and $v_{\rm halo} = 131.5 \, {\rm km\, s^{-1}}$.
The initial density distribution of the satellite is given by a Plummer
profile 
\begin{equation}
\label{eq:density_sat}
\rho(r) =  \frac{\rho_0}{(r^2 + r^2_0)^{5/2}}
\end{equation}
with $\rho_0 = 3 M/4 \pi r_0^3$, $M = 10^{7}\, {\rm M_{\odot}} $ 
being the initial mass of 
the satellite and $r_0 = 0.53 \, {\rm kpc}$ its scale length.
Its one-dimensional internal velocity dispersion is 
$2.9 \, {\rm km \, s^{-1}}$.
We run several simulations which differ in their orbital parameters, which 
span a range in radial periods from 0.5 to 1.3 Gyr, and have 
an apocentre to pericentric distance ratio of 5 to 10 (fairly radial orbits).
We have also imposed that the orbits pass close to the solar circle 
to compare the results of the experiments with the known properties 
of the stellar halo of the Milky Way. 
In all cases the satellite was represented by $10^5$
particles of equal mass.
We find that the satellites become completely unbound after, at most, three
pericentric passages.
 
\subsection{Analysis of the simulations: Structure in phase-space}

One process that will
tend to erase any macroscopic correlation between the particles,
making more difficult the detection of satellite debris, is phase-mixing.
To quantify it we use the coarse-grained entropy defined as:
\begin{equation}
\label{eq:def_entropy}
S[\bar{f}] = -\int \bar{f}\ln \bar{f} d^3x d^3v,
\end{equation}
where $\bar{f}$ is the coarse-grained distribution function, that is, the
average of the actual distribution function $f$ 
over small cells in phase-space.
One of the interesting properties of $\bar{f}$ is that it decreases as
the system becomes phase-mixed. Therefore, $S[f]$ is expected to 
increase with time, as shown for our simulations in Figure~1. 
In practice, we replace the integral by a sum over cells, 
and $\bar{f}$ by the fraction of particles in
each cell.
\begin{figure}[!htb]
\plotfiddle{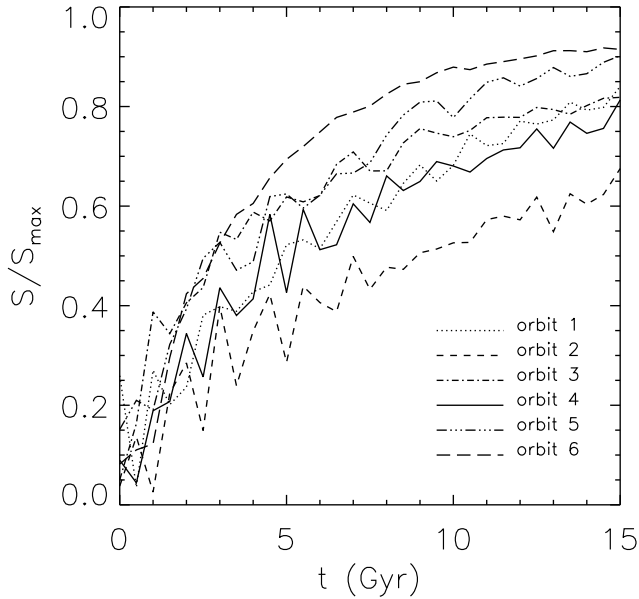}{75truemm}{0}{45}{45}{-150}{-80}
\caption{Evolution of the entropy of the system for the different 
experiments, as a function of time. Orbits which go deeper 
into the potential, and have the shortest periods, show the most 
advanced state of mixing, which is not complete after a Hubble time.}
\end{figure}

The spatial properties of the debris can be studied by plotting isodensity 
surfaces. These surfaces
are indicative of how spread in its available configuration volume 
the system is, or equivalently how advanced the disruption is. 
We find that for the
region of parameter space probed, these volumes are almost completely filled
after a Hubble time. In terms of the spatial distribution of the particles
on the plane of the sky (see Helmi, Zhao and de Zeeuw, this volume, 
their Figure~1) we do not find any strong correlations, contrary to
what Johnston, Hernquist and Bolte (1996) find in their simulations of
accretion in the outer halo. We can understand this in terms of
the short time scales and the strong flattening  
characteristic of the inner parts of our Galaxy. The maximum densities of
such debris are three to four orders of magnitude lower than the   
initial density
of their progenitors, roughly comparable to the local density of the stellar
halo.

In order to reflect what observers can do in surveys of the local halo, 
in Figure~2 we plot the kinematical properties of stars 
inside a box of 3 kpc on a side in different locations along the orbit. 
Notice the strong correlations between the different components 
of the velocity vector inside any given box,
and, in particular, the large velocity range in each component
when close to the Galactic centre. This
shows that the debris can appear kinematically hot. This is the result of
a combination of multiple streams within a
given box (clearly visible in Figure~2) and  
strong gradients along each stream.
At a given point along any particular stream  
the dispersions are usually very small.
\begin{figure*}
\label{fig2}
\plotfiddle{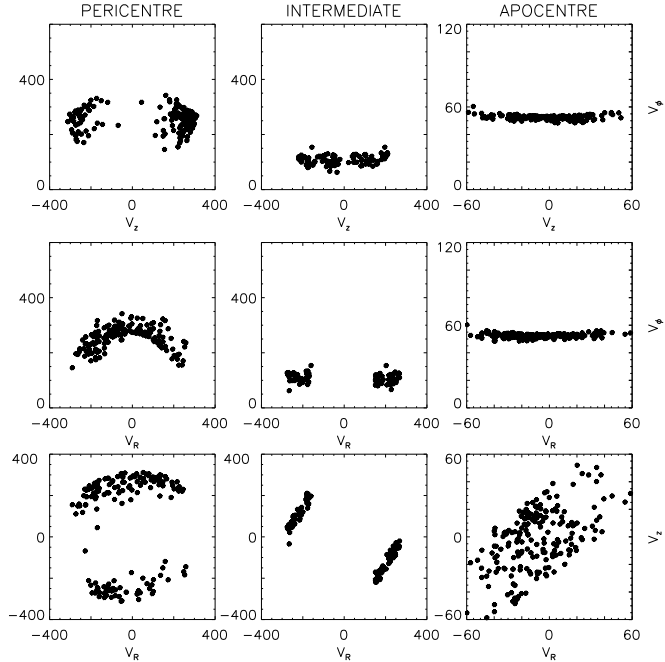}{75truemm}{0}{40}{40}{-150}{-55}
\caption[]{Scatter velocity plots for stars
in boxes of $\sim$ 3 kpc on a side at different locations for one of the
experiments at 13.5 Gyr.}
\end{figure*}

\subsection{What sets the characteristic scales?}

Because the disruption of the satellite occurs very early in its history, 
it can be considered as an ensemble of test 
particles during  most of its evolution. One
of the distinguishing properties of this ensemble is that it initially
had a very high density in phase-space, and by virtue of 
Liouville's theorem, this is true at all times. At later times,
however, this is no longer reflected 
by a strong concentration in configuration space. Since
the behaviour of a dynamical system is particularly simple in action-angle
variables: 
\begin{eqnarray}
\label{eq:evol}
{\mathbf \Phi}\!\!\!& = &\!\!\! {\mathbf \Phi_i} + 
{\mathbf \Omega}({\mathbf{J}})\, t, \nonumber\\
{\mathbf J} \!\!\!& = &\!\!\! {\mathbf J_i} = constant.
\end{eqnarray}
we can find the evolution of the distribution 
function in the following way. We know the distribution function 
at the initial
time $f({\mathbf{x}_i},{\mathbf{v}_i})$. By a transformation of coordinates
we may write it as a function of action-angle variables at that initial
time: $f({\mathbf{\Phi_i}},{\mathbf{J}_i})$. 
Eqs.~(\ref{eq:evol}) then give the distribution function at a later time $t$ in
action-angle variables. If we now transform locally from 
$({\mathbf{\Phi}},{\mathbf{J}})$ back to $({\mathbf{x}},{\mathbf{v}})$, we
obtain the distribution function, and thereby the velocity dispersions
and the density behaviour, in the region of interest.
In action-angle variables the system expands along three
directions and contracts along the remaining three. This is directly 
associated with the equations of motion. Any initial dispersion in
the angles can only increase (Eqs.~(\ref{eq:evol})), so that the
system becomes a very elongated ellipsoid in phase-space as time passes by.
The conservation of phase-space density (Liouville's theorem) then 
forces the
other directions to shrink. This is reflected in the projection
onto observable space. It is possible to show (see Helmi
\& White 1998 for all the details) that the velocity dispersions decrease
on the average with time, 
and so the volume density also decreases with time
as the system expands in the spatial directions. 
For example in the spherical
case the dispersions and the central density behave as
\[\frac{\sigma(v_{\varphi})}{\sigma_i(v_{\varphi})} \propto \frac{t_o}{t}, 
\qquad
\frac{\sigma(v_{r})}{\sigma_i(v_{r})} \propto \frac{t_o}{t}, \qquad 
 \frac{\rho}{\rho_i}\propto \left(\frac{t_o}{t}\right)^{2}. \]
The velocity dispersion in the direction  transverse to the plane
of motion ($\theta$) is constant, 
except for periodic variations due to the orbital 
phase. In the axisymmetric case, there no longer is a preferred
orientation, so that also in the $\theta$-direction the velocity dispersion
decreases, and therefore the density decreases as $t^{-3}$.

\section{Discussion}
Our results suggest that fossil structure
from an accretion event which took place at any time 
during the the history of
the Milky Way should be visible in velocity space.
A stream of stars is, most of the time, fairly cold. 
However, in particular in the inner parts of the Galaxy, it is quite
likely to find more than one stream from any particular disrupted
satellite in a small region in configuration 
space. In that case, the apparent velocity dispersions can be much
larger, although constrained by the initial dispersions in the integrals
of motion. 
Majewski, Munn \& Hawley (1994) reported the discovery of a 
moving group near the
NGP (for details see Majewski, this volume) with large velocity dispersions
(greater than 30 ${\rm {km \,s^{-1}}}$), and with a mean motion very
different from that of the other stars in the field. If we are to take
these stars as satellite
debris, we will have to invoke a multistream structure in order to 
explain their 
kinematics. Indeed, there is some evidence of substructure
in their distribution of angular
momenta. This is to be expected if indeed there are multiple streams. 
With this in mind, we can use our simulations and analytic results to 
estimate the mass 
of the progenitor using the simulations, 
and estimates for its initial size and velocity dispersions. We find: 
$M \sim 10^9 \,{\rm M}_{\odot}$, $R \sim 6 \,{\rm kpc}$ and 
$\sigma(v) \sim 28 \,{\rm {km \, s^{-1}}}$.

\acknowledgments
A. H. wishes to thank Martin Bureau for all the help 
during the meeting, and acknowledges financial support from
LKBF and Zonta International through an Amelia Earhart Fellowship award.

\end{document}